\documentclass[twocolumn,pra,superscriptaddress,showpacs,amsmath]{revtex4}

\usepackage{graphicx}

\begin{document}

\title
     {Light scattering from ultracold atoms in optical lattices as an
      optical probe of quantum statistics}
\date{\today}

\author{Igor B. Mekhov}
\email{Igor.Mekhov@uibk.ac.at} \affiliation{Institut f\"ur
Theoretische Physik, Universit\"at Innsbruck, Innsbruck, Austria}
\affiliation{St. Petersburg State University, Faculty of Physics,
St. Petersburg, Russia}
\author{Christoph Maschler}
\author{Helmut Ritsch}
\affiliation{Institut f\"ur Theoretische Physik, Universit\"at
Innsbruck, Innsbruck, Austria}

\begin{abstract}
We study off-resonant collective light scattering from ultracold
atoms trapped in an optical lattice. Scattering from different
atomic quantum states creates different quantum states of the
scattered light, which can be distinguished by measurements of the
spatial intensity distribution, quadrature variances, photon
statistics, or spectral measurements. In particular, angle-resolved
intensity measurements reflect global statistics of atoms (total
number of radiating atoms) as well as local statistical quantities
(single-site statistics even without an optical access to a single
site) and pair correlations between different sites. As a striking
example we consider scattering from transversally illuminated atoms
into an optical cavity mode. For the Mott insulator state, similar
to classical diffraction, the number of photons scattered into a
cavity is zero due to destructive interference, while for the
superfluid state it is nonzero and proportional to the number of
atoms. Moreover, we demonstrate that light scattering into a
standing-wave cavity has a nontrivial angle dependence, including
the appearance of narrow features at angles, where classical
diffraction predicts zero. The measurement procedure corresponds to
the quantum non-demolition (QND) measurement of various atomic
variables by observing light.

\end{abstract}


\maketitle

\section{Introduction}

Ever since the first generation of Bose-Einstein condensates (BEC),
it has been a central task to study quantum properties of such
degenerate gases. Surprisingly, it turned out that many properties
are well explained by the Gross-Pitaevskii equation, which is a an
effective nonlinear single-particle equation and allows to calculate
the evolution of the average atomic density and phase. The density
can be observed by simple absorption images after expansion, and the
phase can be mapped onto density modulations in interferometric
setups. The limited validity of such mean-field descriptions became
apparent with the advent of optical lattices \cite{jaksch,greiner},
where one has quantum phase transitions between states of similar
average density but radically different quantum fluctuations.

The majority of methods to characterize quantum properties of
degenerate gases are based on matter-wave interference between atoms
released from a trap in time-of-flight measurements
\cite{greiner,blochnew,lukin}, which destroys the system. Recently,
a method of ``Bragg spectroscopy'' based on stimulated scattering of
matter waves by laser pulses was applied to homogeneous BECs
\cite{stenger,blakie} and atoms in lattices
\cite{roth,stoferle,menotti,rey}. In this case, the measured
quantities (e.g. structure factor), which carry information about
density fluctuations, are also accessible via matter-wave
interference. Although the scattered light and stimulated matter
waves can be entangled and mutually carry the statistical
information \cite{moore,pu}, the laser fields are simply considered
as a tool to stimulate matter waves.

In contrast to those works, the nondestructive methods based on
measurements of light fields only, without destroying atoms, were
proposed in Refs.~\cite{you,idziaszek,mustPRA62,mustPRA64,
javPRL,javPRA,ciracPRL,ciracPRA,saito,prataviera} for homogeneous
BECs in traps and optical lattices \cite{javOL}. Here the average
amplitude of the scattered light is solely determined by the average
atomic density, while the photon number and other higher order field
expectation values contain quantum statistical properties of atoms.

In this paper, we show that this is of even greater significance for
atoms in lattices, where different quantum phases show qualitatively
distinct light scattering. Here we extend the preliminary results
presented in our previous letter, Ref.~\cite{PRL}. In particular,
linear scattering can create entangled states of light and manybody
atomic states, exhibiting a nontrivial connection of the field
amplitude and intensity. As a practical consequence, we demonstrate
the possibility to distinguish between different quantum phases,
e.g., Mott insulator (MI) and superfluid (SF), by measuring
properties of a scattered off-resonant beam. This possibility is
exhibited in several different ways involving simple intensity
measurements, or more involved measurements of quadrature variances,
photon statistics, as well as phase-sensitive or spectral
measurements. A careful analysis of the scattered light provides
information about global statistics (related to atom number at a
lattice region illuminated by the probe), local quantities
(reflecting statistics at a single site even without an optical
single-site access), and pair correlations between different sites.

Note that we consider off-resonant and almost nondestructive light
scattering, corresponding to the quantum non-demolition (QND)
measurements of various atom-number functions. In principle, it can
be repeatedly or even continuously applied to the same sample. This
is very different from noise spectroscopy in absorption images
\cite{blochnew} where observations of quantum fluctuations of the
atomic density were recently reported.

For homogeneous BECs \cite{you,idziaszek,mustPRA62,mustPRA64,
javPRL,javPRA,ciracPRL,ciracPRA,saito,prataviera}, the scattered
light was shown to consist of two contributions: the strong
classical part insensitive to atomic fluctuations, and weaker one,
which carries information about atom statistics. For a large atom
number, the classical part completely dominates the second one,
which, in some papers, even led to a conclusion about the
impossibility of distinguishing between different atomic states by
intensity measurements, and, hence, to a necessity to measure photon
statistics.

In our work, we show that light scattering from atoms in optical
lattices has essentially different and advantageous characteristics
in contrast to scattering from homogeneous BECs. For example, the
problem of suppressing the strong classical part of scattering has a
natural solution: in the directions of classical diffraction minima,
the expectation value of the light amplitude is zero, while the
intensity (photon number) is nonzero and therefore directly reflects
density fluctuations. Furthermore, in an optical lattice, the signal
is sensitive not only to the periodic density distribution, but also
to the periodic distribution of density fluctuations, giving an
access to even very small nonlocal pair correlations, which is
possible by measuring light in the directions of diffraction maxima.

As free space light scattering from a small sample can be weak, it
might be selectively enhanced by a cavity. The corresponding light
scattering from an optical lattice exhibits a complicated angle
dependence and narrow angle-resolved features appear at angles,
where classical diffraction cannot exist. In experiments, such a
nontrivial angle dependence can help in the separation between the
signal reflecting atom statistics from a technical noise.

Joining the paradigms of two broad fields of quantum physics, cavity
quantum electrodynamics (QED) and ultracold gases, will enable new
investigations of both light and matter at ultimate quantum levels,
which only recently became experimentally
possible~\cite{EsslingerQED,VuleticQED,Reichel}. Here we predict
effects accessible in such novel setups.

Experimentally, diffraction (Bragg scattering) of light from
classical atoms in optical lattices was considered, for example, in
Refs.~\cite{birkl,weidemuller,slama}. In our work, we are
essentially focused on the properties of scattering from ultracold
lattice atoms with quantized center-of-mass motion.

The paper is organized as follows. In Sec.~II, a general theoretical
model of light scattering from atoms in an optical lattice is
developed taking into account atom tunneling between neighboring
sites. In Sec.~III, we significantly reduce the model to the case of
a deep lattice and give a classical analogy of light diffraction on
a quantum lattice. Section~IV presents a relation between atom
statistics and different characteristics of scattered light:
intensity and amplitude, quadratures, photon statistics, and
phase-sensitive and spectral characteristics. Properties of
different atomic states are summarized in Sec.~V.

In Sec.~VI, we present a simple example of the model developed:
light scattering from a lattice in an optical cavity pumped
orthogonally to the axis. The main results are discussed in Sec.~VII
and summarized in Sec.~VIII.

\section{General model}

\begin{figure}
\scalebox{0.7}[0.7]{\includegraphics{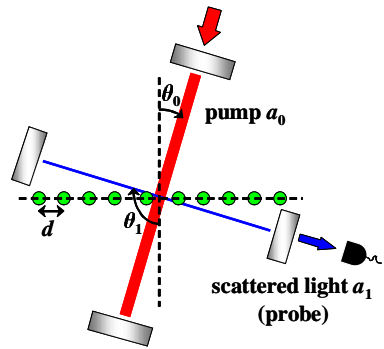}}
\caption{\label{fig1}(Color online) Setup. Atoms in a lattice are
illuminated by a pump wave at angle $\theta_0$; scattered (probe)
light is collected by a cavity at angle $\theta_1$ and measured by a
detector.}
\end{figure}

We consider an ensemble of $N$ two-level atoms in an optical lattice
with $M$ sites. Except the presence of a trapping lattice potential,
atoms are illuminated by and scatter field modes at different
directions. A possible experimental realization is shown in
Fig.~\ref{fig1}. Here, a lattice is illuminated by a ``pump'' beam,
whereas measurements are carried out in one of the scattered modes,
which is treated as a ``probe.'' Note that different experimental
setups are possible: the modes can be either in free space, or
selected by traveling- or standing-wave cavities, or even correspond
to different modes of the same cavity. For definiteness, we will
consider the case, where mode functions are determined by cavities,
whose axes directions can be varied with respect to the lattice axis
(the simplest case of two standing-wave cavities at angles
$\theta_0$ and $\theta_1$ is shown in Fig.~\ref{fig1}). Instead of
varying the angles, the mode wavelengths can be varied with respect
to the wavelength of a trapping beam. We also assume, that not all
$M$ lattice sites are necessarily illuminated by additional modes,
but some region with $K\le M$ sites.

The manybody Hamiltonian in the second quantized form is given by
\begin{subequations}\label{1}
\begin{eqnarray}
H=H_f +H_a, \\
H_f=\sum_l{\hbar\omega_l a^\dag_l a_l} -i\hbar\sum_l{(\eta^*_l a_l -
\eta_l a^\dag_l)}, \\
H_a=\int{d^3{\bf r}\Psi^\dag({\bf r})H_{a1}\Psi({\bf r})} \nonumber\\
+\frac{2\pi a_s \hbar^2}{m}\int{d^3{\bf r}\Psi^\dag({\bf
r})\Psi^\dag({\bf r})\Psi({\bf r})\Psi({\bf r})}.
\end{eqnarray}
\end{subequations}
In the field part of the Hamiltonian $H_f$, $a_l$ are the
annihilation operators of light modes with the frequencies
$\omega_l$, wave vectors ${\bf k}_l$, and mode functions $u_l({\bf
r})$, which can be pumped by coherent fields with amplitudes
$\eta_l$. In the atom part, $H_a$, $\Psi({\bf r})$ is the atomic
matter-field operator, $a_s$ is the $s$-wave scattering length
characterizing the direct interatomic interaction, and $H_{a1}$ is
the atomic part of the single-particle Hamiltonian $H_1$, which in
the rotating-wave and dipole approximation has a form

\begin{subequations}\label{2}
\begin{eqnarray}
H_1=H_f +H_{a1}, \\
H_{a1}=\frac{{\bf p}^2}{2m_a}+\frac{\hbar\omega_a}{2} \sigma_z -
i\hbar g_0\sum_l{[\sigma^+ a_l u_l({\bf r})-\text{H. c.}}]
\end{eqnarray}
\end{subequations}
Here, ${\bf p}$ and ${\bf r}$ are the momentum and position
operators of an atom of mass $m_a$ and resonance frequency
$\omega_a$, $\sigma^+$, $\sigma^-$, and $\sigma_z$ are the raising,
lowering, and population difference operators, $g_0$ is the
atom--light coupling constant.

We will consider essentially nonresonant interaction where the
light-atom detunings $\Delta_{la} = \omega_l - \omega_a$ are much
larger than the spontaneous emission rate and Rabi frequencies $g_0
a_l$. Thus, in the Heisenberg equations obtained from the
single-atom Hamiltonian $H_1$ (\ref{2}), $\sigma_z$ can be set to
$-1$ (approximation of linear dipoles). Moreover, the polarization
$\sigma^-$ can be adiabatically eliminated and expressed via the
fields $a_l$. An effective single-particle Hamiltonian that gives
the corresponding Heisenberg equation for $a_l$ can be written as
$H_{1\text{eff}}=H_f +H_{a1}$ with
\begin{eqnarray}\label{3}
H_{a1}=\frac{{\bf p}^2}{2m_a}+V_{\text {cl}}({\bf r})+\hbar
g^2_0\sum_{l,m}{\frac{u_l^*({\bf r})u_m({\bf r}) a^\dag_l
a_m}{\Delta_{ma}}}.
\end{eqnarray}
Here, we have also added a classical trapping potential of the
lattice, $V_{\text {cl}}({\bf r})$, corresponds to a strong
classical standing wave. This potential can be, of course, derived
from one of the modes $a_l = a_{\text {cl}}$ [in this case $V_{\text
{cl}}({\bf r})=\hbar g^2_0|a_{\text {cl}} u_{\text {cl}}({\bf
r})|^2/\Delta_{\text {cl}a}$], and it can scatter light into other
modes. Nevertheless, at this point we will consider $V_{\text
{cl}}({\bf r})$ as an independent potential, which does not affect
light scattering of other modes that can be significantly detuned
from $a_{\text {cl}}$ [i.e. the interference terms between $a_{\text
{cl}}$ and other modes are not considered in the last term of
Eq.~(\ref{3})]. The later inclusion of the light scattered by the
trapping wave will not constitute a difficulty, due to the linearity
of dipoles assumed in this model.

We will consider scattering of weak modes from the atoms in a deep
lattice. So, the fields $a_l$ are assumed much weaker than the field
forming the potential $V_{\text {cl}}({\bf r})$. To derive the
generalized Bose--Hubbard Hamiltonian we expand the field operator
$\Psi({\bf r})$ in Eq.~(\ref{1}), using localized Wannier functions
corresponding to $V_{\text {cl}}({\bf r})$ and keeping only the
lowest vibrational state at each site: $\Psi({\bf
r})=\sum_{i=1}^{M}{b_i w({\bf r}-{\bf r}_i)}$, where $b_i$ is the
annihilation operator of an atom at the site $i$ with the coordinate
${\bf r}_i$. Substituting this expansion in Eq.~(\ref{1}) with
$H_{a1}$ (\ref{3}), we get
\begin{eqnarray}\label{4}
H=H_f+\sum_{i,j=1}^M{J_{i,j}^{\text {cl}}b_i^\dag b_j} + \hbar g^2_0
\sum_{l,m}{\frac{a^\dag_l
a_m}{\Delta_{ma}}}\left(\sum_{i,j=1}^K{J_{i,j}^{lm}b_i^\dag
b_j}\right)  \nonumber \\
+\frac{U}{2}\sum_{i=1}^M{b_i^\dag b_i(b_i^\dag b_i-1)},
\end{eqnarray}
where the coefficients $J_{ij}^{\text {cl}}$ correspond to the
quantum motion of atoms in the classical potential and are typical
for the Bose--Hubbard Hamiltonian \cite{jaksch}:

\begin{equation}\label{5}
J_{i,j}^{\text {cl}}=\int{d{\bf r}}w({\bf r}-{\bf
r}_i)\left(-\frac{\hbar^2\nabla^2}{2m}+V_{\text {cl}}({\bf
r})\right)w({\bf r}-{\bf r}_j).
\end{equation}
However, in contrast to the usual Bose--Hubbard model, one has new
terms depending on the coefficients $J_{ij}^{lm}$, which describe an
additional contribution arising from the presence of light modes:
\begin{equation}\label{6}
J_{i,j}^{lm}=\int{d{\bf r}}w({\bf r}-{\bf r}_i) u_l^*({\bf
r})u_m({\bf r})  w({\bf r}-{\bf r}_j).
\end{equation}
In the last term of Eq.~(\ref{4}), only the on-site interaction was
taken into account and $U=4\pi a_s\hbar^2/m_a \int{d{\bf r}|w({\bf
r})|^4}$.

As a usual approximation, we consider atom tunneling being possible
only to the nearest neighbor sites. Thus, coefficients (\ref{5}) do
not depend on the site indices ($J_{i,i}^{\text {cl}}=J_0^{\text
{cl}}$ and $J_{i,i\pm 1}^{\text {cl}}=J^{\text {cl}}$), while
coefficients (\ref{6}) are still index-dependent. The Hamiltonian
(\ref{4}) then reads
\begin{eqnarray}\label{7}
H=H_f+J_0^{\text {cl}}\hat{N}+J^{\text {cl}}\hat{B}+ \hbar g^2_0
\sum_{l,m}{\frac{a^\dag_l
a_m}{\Delta_{ma}}}\left(\sum_{i=1}^K{J_{i,i}^{lm}\hat{n}_i}\right)  \nonumber \\
+\hbar g^2_0 \sum_{l,m}{\frac{a^\dag_l
a_m}{\Delta_{ma}}}\left(\sum_{<i,j>}^K{J_{i,j}^{lm}b_i^\dag
b_j}\right) +\frac{U}{2}\sum_{i=1}^M{\hat{n}_i(\hat{n}_i-1)},
\end{eqnarray}
where $<i,j>$ denotes the sum over neighboring pairs,
$\hat{n}_i=b_i^\dag b_i$ is the atom number operator at the $i$-th
site, and $\hat{B}=\sum_{i=1}^M{b^\dag_i b_{i+1}}+{\text {H.c.}}$
While the total atom number determined by
$\hat{N}=\sum_{i=1}^M{\hat{n}_i}$ is conserved, the atom number at
the illuminated sites, determined by
$\hat{N}_K=\sum_{i=1}^K{\hat{n}_i}$, is not necessarily a conserved
quantity.

The Heisenberg equations for $a_l$ and $b_i$ can be obtained from
the Hamiltonian (\ref{7}) as
\begin{subequations}\label{8}
\begin{eqnarray}
\dot{a}_l= -i\left( \omega_l
+\frac{g_0^2}{\Delta_{la}}\sum_{i=1}^K{J_{i,i}^{ll}\hat{n}_i}+
\frac{g_0^2}{\Delta_{la}}\sum_{<i,j>}^K{J_{i,j}^{ll}b_i^\dag
b_j}\right) a_l    \nonumber \\
-ig_0^2\sum_{m \ne
l}\frac{a_m}{\Delta_{ma}}\left(\sum_{i=1}^K{J_{i,i}^{lm}\hat{n}_i}\right) \nonumber \\
-ig_0^2\sum_{m \ne
l}\frac{a_m}{\Delta_{ma}}\left(\sum_{<i,j>}^K{J_{i,j}^{lm}b_i^\dag
b_j}\right)+ \eta_l  \label{8a}    \\
\dot{b}_i=-\frac{i}{\hbar}\left( J_0^{\text {cl}}+\hbar
g_0^2\sum_{l,m}{\frac{a^\dag_l
a_m}{\Delta_{ma}}J_{i,i}^{lm}}+U\hat{n}_i\right) b_i  \nonumber \\
-\frac{i}{\hbar}\left( J^{\text {cl}}+\hbar
g_0^2\sum_{l,m}{\frac{a^\dag_l
a_m}{\Delta_{ma}}J_{i,i+1}^{lm}}\right)b_{i+1} \nonumber \\
-\frac{i}{\hbar}\left( J^{\text {cl}}+\hbar
g_0^2\sum_{l,m}{\frac{a^\dag_l
a_m}{\Delta_{ma}}J_{i,i-1}^{lm}}\right)b_{i-1}. \label{8b}
\end{eqnarray}
\end{subequations}

In Eq.~(\ref{8a}) for the electromagnetic fields $a_l$, two last
terms in the parentheses correspond to the phase shift of the light
mode due to nonresonant dispersion (the second term) and due to
tunneling to neighboring sites (the third one). The second term in
Eq.~(\ref{8a}) describes scattering of all modes into $a_l$, while
the forth term takes into account corrections to such scattering
associated with tunneling due to the presence of additional light
fields. In Eq.~(\ref{8b}) for the matter field operators $b_i$, the
first term gives the phase of the matter-field at the site $i$, the
second and third terms describe the coupling to neighboring sites.

It is important to underline that except for the direct coupling
between neighboring sites, which is usual for the standard
Bose--Hubbard model, Eqs.~(\ref{8}) also take into account
long-range correlations between sites, which do not decrease with
the distance and are provided by the common light modes $a_l$ that
are determined by the whole set of matter-field operators $b_i$.
Such nonlocal correlations between the operators $b_i$, which are
introduced by the general Eqs.~(\ref{8}), can give rise to new
many-body effects beyond predictions of the standard Bose-Hubbard
model \cite{maschler}.

\section{Scattering from a deep lattice and classical analogy}

We will significantly reduce the general model described by the
Hamiltonian (\ref{7}) and Heisenberg equations (\ref{8}). In
contrast to our paper \cite{maschler} and works on so-called ``Bragg
spectroscopy'' \cite{roth,stoferle,menotti,rey}, we will not
consider excitations of the lattice by light and stimulation of
matter waves. The focus of the present paper is a study of
properties of light scattered from the atoms in a prescribed quantum
state, which is not necessarily the ground one. The main result is
the demonstration of the possibility to distinguish between
different atomic quantum states of different statistics by measuring
light only.

We consider a deep lattice formed by a strong classical potential
$V_{\text {cl}}({\bf r})$, so that the overlap between Wannier
functions in Eqs.~(\ref{5}) and (\ref{6}) is small. Thus, we can
neglect the contribution of tunneling to the scattered light by
putting $J^{\text{cl}}=0$ and $J_{i,j}^{lm}=0$ for $i \ne j$. Under
this approximation, the matter-wave dynamics is not essential for
light scattering. In experiments, such situation can be realized
because the time scale of light measurements can be much faster than
the time scale of atomic tunneling. One of the well-known advantages
of the optical lattices is their extremely high tunability. Thus,
tuning the lattice potential, tunneling can be made very
slow~\cite{jaksch}. On the other hand, the rate of the photon escape
from the cavity depends on the cavity relaxation rate and photon
number, while the letter is determined by various parameters as
atom-field detuning, pump amplitude, and atom number. In modern
experimental setups, all of these parameters, especially the
ultracold atom number, can be tuned in a very broad range. Moreover,
even with no tunneling, various atomic quantum states with
essentially different statistics can be realized until the atoms
will decay to the ground state. Since in this paper we do not
require the atoms to be in a ground state, light scattering can
reflect different atomic statistics even for negligible tunneling.

In a deep lattice, the on-site coefficients $J_{i,i}^{lm}$ (\ref{6})
can be approximated as $J_{i,i}^{lm}=u_l^*({\bf r}_i)u_m({\bf r}_i)$
neglecting details of the atomic localization. Such details are
accessible even from the classical consideration
\cite{birkl,weidemuller,slama}. In this paper, we will focus on
essentially quantum aspects of the problem.

For simplicity, we will consider scattering of a single mode $a_0$
(``pump''), considered as a given operator, into another mode $a_1$
with the relaxation rate $\kappa$ included phenomenologically. The
Heisenberg equation (\ref{8a}) for the scattered light $a_1$ then
reads

\begin{eqnarray}\label{9}
\dot{a}_1= -i\left[ \omega_1
+\frac{g_0^2}{\Delta_{1a}}\sum_{i=1}^K{|u_1({\bf
r}_i)|^2\hat{n}_i}\right]a_1  \nonumber\\
-i\frac{g_0^2 a_0}{\Delta_{0a}}\left[\sum_{i=1}^K{u_1^*({\bf
r}_i)u_0({\bf r}_i)\hat{n}_i}\right]-\kappa a_1 +\eta_1,
\end{eqnarray}
where we do not add the Langevin noise term, since we will be
interested in normal ordered quantities only. In the Heisenberg
equation for the matter--field operators $b_i$ (\ref{8b}), only the
first term is nonzero. This term affects only the phase of the
matter field, but not the atom number operators $\hat{n}_i=b^\dag_i
b_i$. Hence, though the the matter--field phase still depends on the
common light mode, the operators $\hat{n}_i$, appearing in
Eq.~(\ref{9}), are constant in time.

We assume that the dispersion shift of the cavity mode
$g_0^2/\Delta_{1a}\sum_{i=1}^K{|u_1({\bf r}_i)|^2\hat{n}_i}$ is much
smaller than $\kappa$ or detuning between the pump and scattered
light $\Delta_{01}=\omega_0-\omega_1$. Thus, a stationary solution
of Eq.~(\ref{9}) has a form
\begin{subequations}\label{10}
\begin{eqnarray}
a_1=C\hat{D}, \qquad C\equiv -\frac{ig_0^2
a_0}{\Delta_{0a}(\kappa-i\Delta_{01})},   \label{10a}\\
\hat{D}\equiv \sum_{i=1}^K{u_1^*({\bf r}_i)u_0({\bf r}_i)\hat{n}_i},
\label{10b}
\end{eqnarray}
\end{subequations}
where we have also assumed no additional pumping ($\eta_1 =0$) and
replaced the operators $a_{0,1}(t)$ by their slowly varying
envelopes $\tilde{a}_{0,1}(t)$
[$a_{0,1}(t)=\tilde{a}_{0,1}\exp(-i\omega_0 t)$] skipping in the
following notations all tilde signs.

Expressing the light operators in terms of the atomic ones in
Eq.~(\ref{10}) is a central result here, which we will use to study
the properties of the scattered field. The dependence of the light
Heisenberg operators on the atomic operators reflects the
entanglement between light and matter during the light-matter
interaction. We will assume the pump mode to be in the coherent
state, which enables us to consider the quantity $a_0$ as a
c-number.

In the following, we will consider a 1D lattice of the period $d$
with atoms trapped at $x_m=md$ ($m=1,2,..,M$). The result for the
field operator $a_1$ (\ref{10a}) with the operator $\hat{D}$
(\ref{10b}) has an analogy in classical diffraction. For scattering
of a traveling wave $a_0$ in the direction of a traveling wave $a_1$
from a lattice with $\langle \hat{n}_i\rangle =n$ at each site, the
expectation value of the field is given by
\begin{eqnarray}\label{11}
\langle a_1\rangle =C\langle \hat{D}\rangle
=C\sum_{m=1}^K{e^{im\delta k_x d}\langle \hat{n}_m\rangle}
\nonumber\\
=Cn e^{i(K+1)\alpha_-/2}
\frac{\sin{(K\alpha_-/2})}{\sin{(\alpha_-/2})},
\end{eqnarray}
where $\alpha_-=\delta k_x d$, and $\delta k_x =({\bf k}_0-{\bf
k}_1)_x=k(\sin\theta_0 - \sin\theta_1)$ is the projection of the
difference between two wave vectors on the lattice direction,
$\theta_{0,1}$ are the angles between wave vectors and a vector
normal to the lattice direction (cf. Fig.~\ref{fig1}), $k=\omega/c$
for $\omega_0=\omega_1=\omega$.

Equation (\ref{11}) simply describes classical diffraction of the
traveling wave $a_0$ on a diffraction grating formed by equally
spaced atoms with positions of diffraction maxima and minima (i.e.
scattering angles $\theta_1$) determined by the parameter $\alpha_-$
depending on the geometry of incident and scattered waves and
diffraction grating through $\theta_{0}$, $|{\bf k}_{0,1}|$, and
$d$. A more general form of the operator $\hat{D}$ given by
Eq.~(\ref{10b}) describes also diffraction of a standing wave $a_0$
into another mode $a_1$, which can be formed, for example, by a
standing--wave or ring optical cavity.

Equation (\ref{11}) shows that the expectation value of the
scattered field is sensitive only to the mean number of atoms per
site $n$ and reflects a direct analogy of light scattering from a
classical diffraction grating. Nevertheless, the photon number
(intensity) and photon statistics of the field $a_1$ are sensitive
to higher moments of the number operators $\hat{n}_i$ as well as to
the quantum correlations between different lattice sites, which
determines quantum statistical properties of ultracold atoms in an
optical lattice and will be considered in the next sections.

\section{Relation between quantum statistics of atoms and characteristics
of scattered light}

\subsection{Probing quantum statistics by intensity and amplitude measurements}

According to Eq.~(\ref{10a}), the expectation value of the photon
number $a_1^\dag a_1$ is proportional to the expectation value of
the operator $\hat{D}^*\hat{D}$. We introduce coefficients
$A_i(\theta_0,\theta_1)$ responsible for the geometry of the
problem:
\begin{eqnarray}\label{12}
\hat{D}=\sum_{i=1}^K{A_i \hat{n}_i}, \quad
A_i(\theta_0,\theta_1)\equiv u_1^*({\bf r}_i)u_0({\bf r}_i),
\nonumber\\
A(\theta_0,\theta_1)\equiv \sum_{i=1}^K{A_i(\theta_0,\theta_1)},
\end{eqnarray}
where  $u_{0,1}({\bf r}_m)=\exp (imk_{0,1x}d+\phi_{0,1m})$ for
traveling waves, and $u_{0,1}({\bf r}_m)=\cos (mk_{0,1x}d
+\phi_{0,1m})$ for standing waves ($m=1,2,...M$), $k_{0,1x}=|{\bf
k}_{0,1}|\sin\theta_{0,1}$, $\theta_{0,1}$ are the angles between
mode wave vectors and a vector normal to the lattice axis; in the
plane-wave approximation, additional phases $\phi_{0,1m}$ are
$m$-independent.

The expectation values of $\hat{D}$ and $\hat{D}^*\hat{D}$ then read
\begin{subequations}\label{13}
\begin{eqnarray}
\langle \hat{D} \rangle =\sum_{i=1}^K{A_i\langle\hat{n}_i}\rangle= nA, \label{13a}\\
\langle \hat{D}^*\hat{D} \rangle = \sum_{i,j=1}^K{A_i^* A_j
\langle\hat{n}_i\hat{n}_j\rangle}   \label{13b}\\
=\langle \hat{n}_a\hat{n}_b\rangle |A|^2+(\langle\hat{n}^2\rangle -
\langle \hat{n}_a\hat{n}_b\rangle)\sum_{i=1}^K{|A_i|^2}, \label{13c}\\
R(\theta_0, \theta_1)\equiv \langle\hat{D}^*\hat{D} \rangle - |\langle \hat{D} \rangle|^2 =  \nonumber\\
(\langle \hat{n}_a\hat{n}_b\rangle - n^2)
|A|^2+(\langle\hat{n}^2\rangle - \langle
\hat{n}_a\hat{n}_b\rangle)\sum_{i=1}^K{|A_i|^2}  \label{13d}\\
=\langle \delta\hat{n}_a\delta\hat{n}_b\rangle
|A|^2+(\langle\delta\hat{n}^2\rangle - \langle
\delta\hat{n}_a\delta\hat{n}_b\rangle)\sum_{i=1}^K{|A_i|^2}.\label{13e}
\end{eqnarray}
\end{subequations}
In Eqs.~(\ref{13}) we have used the following assumptions about the
atomic quantum state $|\Psi\rangle$: (i) the expectation values of
the atom number at all sites are the same, $\langle\hat{n}_i\rangle
= n$ (thus, the expectation value of atom number at $K$ sites is
$\langle\hat{N}_K\rangle=N_K\equiv nK$), (ii) the nonlocal pair
correlations between atom numbers at different sites
$\langle\hat{n}_i\hat{n}_j\rangle$ are equal to each other for any
$i\ne j$ and will be denoted as $\langle\hat{n}_a\hat{n}_b\rangle$
(with $a\ne b$). The latter assumption is valid for a deep lattice.
We also introduced the fluctuation operators
$\delta\hat{n}_i=\hat{n}_i - n$, which gives
$\langle\delta\hat{n}^2\rangle$ equal to the variance $(\Delta
n_i)^2=\langle\hat{n}_i^2\rangle-n^2$.

Equation (\ref{13a}) reflects the fact that the expectation value of
the field amplitude (\ref{10a}) is sensitive only to the mean atom
numbers and displays the angle dependence of classical diffraction
given by the factor $A(\theta_0,\theta_1)$, which depends on the
mode angles and displays pronounced diffraction maxima and minima.
Equation (\ref{13b}) shows that the number of scattered photons
(intensity) at some angle is determined by the density--density
correlations. In the simplest case of two traveling waves, the
prefactors $A_i^*A_j=\exp[{i\delta k_x(x_j-x_i)]}$ with $\delta k_x
=k_{0x}-k_{1x}$. In this case, Eq.~(\ref{13b}) gives the so-called
structure factor (function), which was considered in the works on
light scattering from homogeneous BEC \cite{javPRL,javPRA}. Here we
essentially focus on optical lattices. Moreover, it will be shown,
that the more general Eq.~(\ref{13b}), which includes scattering of
standing waves, contains new measurable features different from
those of a usual structure factor.

Equation (\ref{13c}) shows, that the angle dependence of the
scattered intensity consists of two contributions. The first term
has an angle dependence $|A(\theta_0,\theta_1)|^2$ identical to that
of the expectation value of the field amplitude squared (\ref{13a}).
The second term is proportional to the quantity
$\langle\hat{n}^2\rangle - \langle \hat{n}_a\hat{n}_b\rangle$ giving
quantum fluctuations and has a completely different angle dependence
$\sum_{i=1}^K{|A_i|^2}$. The expression (\ref{13c}) has a form
similar to the one considered in papers
\cite{you,mustPRA62,mustPRA64} on light scattering from a
homogeneous BEC, where the scattered intensity consisted of two
parts: ``coherent'' (i.e. depending on the average density) and
``incoherent'' one (i.e. depending on the density fluctuations).
Nevertheless, in the present case of a periodic lattice, this
similarity would be exact only in a particular case where there are
no nonlocal pair correlations $\langle \hat{n}_a\hat{n}_b\rangle =
n_a n_b = n^2$ ($\langle \delta\hat{n}_a\delta\hat{n}_b\rangle =0$),
which in general is not true and leads to observable difference
between states with and without pair correlations.

Further insight into a physical role of nonlocal pair correlations
can be obtained from Eqs.~(\ref{13d}) and (\ref{13e}) for the
``noise quantity'' $R(\theta_0, \theta_1)\equiv
\langle\hat{D}^*\hat{D} \rangle - |\langle \hat{D} \rangle|^2$,
where we have subtracted the classical (averaged) contribution
$|\langle \hat{D} \rangle|^2$ to the intensity
$\langle\hat{D}^*\hat{D} \rangle$. Equation (\ref{13e}) shows that,
in the noise quantity, a term with the classical angular
distribution $|A(\theta_0,\theta_1)|^2$ appears only if the pair
correlations are nonzero. The physical meaning of this result is
that, in an optical lattice, it is not only the density distribution
that displays spatial periodic structure leading to diffraction
scattering, but also the distribution of number fluctuations
themselves. In the framework of our assumption about equal pair
correlations, the spatial distribution of fluctuations $\langle
\delta\hat{n}_a\delta\hat{n}_b\rangle$ can be either the same as the
density distribution (with a lattice period $d$) or zero. In the
former case, pair correlations contribute to the first term in
Eqs.~(\ref{13d}) and (\ref{13e}) with classical distribution
$|A(\theta_0,\theta_1)|^2$, in the latter case, $\langle
\delta\hat{n}_a\delta\hat{n}_b\rangle = 0$, and the only signal in
the noise quantity is due to on-site fluctuations
$\langle\delta\hat{n}^2\rangle$ with a different angle dependence
$\sum_{i=1}^K{|A_i|^2}$. Note that, in general, the spatial
distribution of fluctuations can be different from that of the
average density and can have a period proportional to the lattice
period $d$. This will lead to additional peaks in the angular
distribution of the noise quantity (\ref{13d}),~(\ref{13e}). The
generalization of those formulas is straightforward.

Even with spatially incoherent pump $a_0$, the intensity of the
scattered mode $a_1^\dag a_1$ is sensitive to the on-site atom
statistics. To model this situation, the quantum expectation value
$\langle\hat{D}^*\hat{D}\rangle$ (\ref{13b}) should be additionally
averaged over random phases $\phi_{0,1m}$ appearing in the
definition of mode functions in Eq.~(\ref{12}). In Eq.~(\ref{13b}),
only terms with $i=j$ will then survive and the final result reads
\begin{eqnarray}\label{14}
\langle \hat{D}^*\hat{D} \rangle_\text{inc} =p_0 K \langle
\hat{n}^2\rangle ,
\end{eqnarray}
where $p_0$ is equal to 1 for two traveling waves, 1/2 for a
configuration with one standing wave, and 1/4, when both modes
$a_{0,1}$ are standing waves.

\subsection{Quadrature measurements}

The photon number $a_1^\dag a_1$ is determined by the expectation
value $\langle\hat{D}^*\hat{D} \rangle$, whereas $\langle \hat{D}
\rangle$ gives the field $\langle a_1 \rangle$ (\ref{10a}). While
photon numbers can be directly measured, a field $\langle a_1
\rangle$ measurement requires a homodyne scheme. Such a measurement
then makes $\langle \hat{D} \rangle$ experimentally accessible.
Actually for a quantum field only the expectation values of
quadratures of $a_1$ that are Hermitian operators and can be
measured. Using Eq.~(\ref{10a}) and the commutation relation
$[a_1,a_1^\dag]=1$, the quadrature operator $X_\phi$ and its
variance $(\Delta X_\phi)^2$ can be written as
\begin{subequations}\label{15}
\begin{eqnarray}
X_\phi \equiv \frac{1}{2}\left(a_1 e^{-i\phi}+a_1^\dag
e^{i\phi}\right)=|C|\hat{X}^D_{\phi-\phi_C},  \label{15a}\\
X_\phi^2 =\frac{1}{4}+|C|^2 (\hat{X}^D_{\phi-\phi_C})^2, \label{15b}\\
(\Delta X_\phi)^2 \equiv \langle X_\phi^2\rangle - \langle
X_\phi\rangle^2 =\frac{1}{4}+|C|^2 (\Delta X^D_{\phi-\phi_C})^2,
\label{15c}
\end{eqnarray}
\end{subequations}
where $C=|C|\exp({i\phi_C})$ and the quadratures of $\hat{D}$ are
\begin{subequations}\label{16}
\begin{eqnarray}
\hat{X}_\beta^D \equiv \frac{1}{2}\left(\hat{D}
e^{-i\beta}+\hat{D}^*e^{i\beta}\right), \label{16a}\\ (\Delta
X_\beta^D)^2 \equiv \langle (\hat{X}_\beta^D)^2\rangle - \langle
\hat{X}_\beta^D\rangle^2.\label{16b}
\end{eqnarray}
\end{subequations}
In Eqs.~(\ref{15}), the phase $\phi$ is related to the homodyne
reference phase, while $\phi_C$ is determined by the phase of the
pump $a_0$ and parameters of the field--matter system [cf.
Eq.~(\ref{10a})]. Hence, the phase $\beta = \phi-\phi_C$ entering
Eqs.~(\ref{15}) can be controlled by varying the phase difference
between the pump and homodyne fields.

Using Eq.~(\ref{12}), the quadrature operator $\hat{X}_\beta^D$
reads
\begin{eqnarray}\label{17}
\hat{X}_\beta^D =\sum_{i=1}^K{A_i^\beta \hat{n}_i}, \quad
A_i^\beta(\theta_0,\theta_1)\equiv |A_i|\cos{(\phi_{A_i}-\beta)},
\nonumber\\
A^\beta (\theta_0,\theta_1)\equiv
\sum_{i=1}^K{A_i^\beta(\theta_0,\theta_1)},
\end{eqnarray}
where $A_i=|A_i|\exp (i\phi_{A_i})$, and we defined new quantities
$A_i^\beta(\theta_0,\theta_1)$ and $A^\beta(\theta_0,\theta_1)$.

Since Eq.~(\ref{17}) for $\hat{X}_\beta^D$ and Eq.~(\ref{12}) for
$\hat{D}$ have a similar structure, the Eqs.~(\ref{13}) for the
quantities $\langle \hat{D}\rangle$,
$\langle\hat{D}^*\hat{D}\rangle$, and $R$ can be rewritten for the
quantities $\langle\hat{X}_\beta^D\rangle$, $\langle
(\hat{X}_\beta^D)^2\rangle$, and $(\Delta X_\beta^D)^2$,
respectively, with the change of parameters $A_i(\theta_0,\theta_1)$
and $A(\theta_0,\theta_1)$ to $A_i^\beta(\theta_0,\theta_1)$ and
$A^\beta (\theta_0,\theta_1)$. Thus, the above discussion of
Eqs.~(\ref{13}) can be repeated in terms of the quadrature operators
with the only difference that coefficients
$A_i^\beta(\theta_0,\theta_1)$ and $A^\beta (\theta_0,\theta_1)$ now
depend also on the homodyne phase. An advantage of this
reformulation is that the expectation value of the non-Hermitian
operator $a_1$, which determines $\langle \hat{D}\rangle$, is now
replaced by the expectation value of the Hermitian operator
$X_\phi$, which is consistent with a procedure of measuring
quadratures of the quantum field $a_1$. The well-known relations
between intracavity and outcoupled fields can be found, e.g., in
Ref.~\cite{Walls} for linear systems, which is the case in our work.

\subsection{Photon number fluctuations}

While the intensity of the scattered light is sensitive to the
second moments of the number operators $\hat{n}_i$, quantum
statistics of the field reflexes the forth-order moments. The
variance $(\Delta n_\text{ph})^2$ of the photon number
$n_\text{ph}=a_1^\dag a_1$ is given by
\begin{eqnarray}\label{18}
(\Delta n_\text{ph})^2=\langle n_\text{ph}^2\rangle - \langle
n_\text{ph}\rangle^2
=:(\Delta n_\text{ph}^2): +\langle n_\text{ph}\rangle \nonumber\\
= |C|^4(\langle \hat{D}^{*2}\hat{D}^2\rangle -\langle
\hat{D}^*\hat{D}\rangle^2)+|C|^2\langle \hat{D}^*\hat{D}\rangle,
\end{eqnarray}
where $:(\Delta n_\text{ph}^2): = \langle a_1^{\dag 2}a_1^2 \rangle
- \langle a_1^{\dag}a_1\rangle^2=|C|^4(\langle
\hat{D}^{*2}\hat{D}^2\rangle -\langle \hat{D}^*\hat{D}\rangle^2)$ is
a normal ordered photon-number variance. Thus, the problem is
reduced to measurements of the photon number $|C|^2\langle
\hat{D}^*\hat{D} \rangle$ and quantity $|C|^4\langle
\hat{D}^{*2}\hat{D}^2\rangle$, which after straightforward
calculations is given by
\begin{widetext}
\begin{eqnarray}\label{19}
\langle \hat{D}^{*2}\hat{D}^2\rangle =\left|\sum_{i=1}^K
A_i\right|^4\langle n_an_bn_cn_d\rangle +2\left[\left(\sum_{i=1}^K
|A_i|^2A_i\right)\sum_{i=1}^K A_i^*+ \text{c.c.}\right](2\langle
n_an_bn_cn_d\rangle-3\langle n_a^2n_bn_c\rangle+\langle
n_a^3n_b\rangle) \nonumber\\
+\left[\left(\sum_{i=1}^K A_i^2\right)\left(\sum_{i=1}^K
A_i^*\right)^2+ \text{c.c.}\right](-\langle
n_an_bn_cn_d\rangle+\langle n_a^2n_bn_c\rangle)
+2\left(\sum_{i=1}^K|A_i|^2\right)^2(\langle
n_an_bn_cn_d\rangle-2\langle n_a^2n_bn_c\rangle+\langle
n_a^2n_b^2\rangle) \nonumber\\
+\left|\sum_{i=1}^K A_i^2\right|^2(\langle
n_an_bn_cn_d\rangle-2\langle n_a^2n_bn_c\rangle+\langle
n_a^2n_b^2\rangle) +4\left|\sum_{i=1}^K A_i\right|^2\sum_{i=1}^K
|A_i|^2(-\langle n_an_bn_cn_d\rangle+\langle n_a^2n_bn_c\rangle) \nonumber\\
+\sum_{i=1}^K |A_i|^4(-6\langle n_an_bn_cn_d\rangle+12\langle
n_a^2n_bn_c\rangle-4\langle n_a^3n_b\rangle-3\langle
n_a^2n_b^2\rangle+\langle n^4\rangle),\quad
\end{eqnarray}
\end{widetext}
where we assumed again that correlations do not depend on site
indices, and sites with the indices $a$, $b$, $c$, and $d$ are
different. In Eq.~(\ref{19}), each prefactor containing geometrical
coefficients $A_i$ (\ref{12}) determines different angle dependences
of a corresponding term.

Thus, varying the geometry of a problem (e.g. angles of two modes,
wavelengths of the modes or that of trapping potential determining
the lattice period), one has access to different statistical
quantities characterizing the quantum state of ultracold atoms.

\subsection{Phase-sensitive and spectral measurements}

In the derivation of Eq.~(\ref{10}), we have neglected the term
$g_0^2/\Delta_{1a}\sum_{i=1}^K{|u_1({\bf r}_i)|^2\hat{n}_i}$ in
Eq.~(\ref{9}) related to the refractive index of atoms for the
scattered light. This term is normally very small at large
detunings. However, if the scattered mode is confined in a very good
optical resonator, the light experiences a very long effective path
within the atoms, and this term shifts the phase of the scattered
light. In a steady state approximation it amounts to the dispersion
shift of a cavity mode.

Equation (\ref{9}) shows that even in the absence of the pump field
($a_0=0$), quantum fluctuations of the atom number enter the phase
via operators $\hat{n}_i$ of $K$ illuminated sites, which depend on
the atomic quantum state. In the simplest case of a traveling wave,
$|u_1({\bf r}_i)|=1$, and the operator
$\sum_{i=1}^K{\hat{n}_i}=\hat{N}_K$ is the number of atoms at $K$
sites. As will be discussed below, in the Mott insulator state, the
expectation value of this quantity $N_K=\langle \hat{N}_K\rangle$
does not fluctuate. In the superfluid state with $K=M$, $N_M$ is
equal to the total number of atoms $N$, and also is fixed. However
for $K<M$, $N_K$ fluctuates strongly and, as will be discussed, for
$K\ll M$ corresponds to a coherent state with $\langle
\hat{N}_K^2\rangle =\langle \hat{N}_K\rangle^2+\langle
\hat{N}_K\rangle$. Those statistical properties of the atomic states
are reflected in the phase of the light field. In particular,
measurements of the dispersion shift of a cavity mode will show a
frequency distribution reflecting the distribution of atom numbers.

This also opens an alternative spectral method of determining the
quantum state of the atoms in a cavity with two degenerate modes.
Let us consider the mode $a_0$ as a dynamical quantity obeying an
equation as Eq.~(\ref{9}), which can be obtained from the set of
Eq.~(\ref{8a}), while the second degenerate mode is called $a_1$.
The atoms lead to the collective normal-mode splitting of two cavity
modes as recently experimentally observed \cite{klinner}. If the
coupling coefficient between two degenerate modes, which is equal to
$g_0^2/\Delta_{0a}\hat{D}$ [cf. Eq.~(\ref{9}) and the definition of
the operator $\hat{D}$ in Eq.~(\ref{12})], exceeds the cavity
relaxation rate $\kappa$, a spectral doublet instead of single
maximum can be observed in the spectrum of the output light.

It is quite expected that the collective strong coupling between the
modes and thus the spectral splitting depends on the number of atoms
in a lattice. Interestingly, from the equations for $a_0$ and $a_1$,
it can be shown, that parameters of the normal-mode splitting (e.g.
splitting frequency, linewidths) also depend on the atomic quantum
state. So, spectral mode-splitting measurements also can be used to
distinguish between atomic quantum phases and allow a nondestructive
measurement of a quantum phase transition dynamics. In the following
we will, however, restrict our study to single frequency
measurements and leave a more detailed analysis of phase- and
frequency-sensitive phenomena to other works~\cite{WeArxiv07}.

\section{Quantum statistical properties of typical atomic distributions}

Let us briefly summarize some key statistical properties of typical
states of $N$ atoms at $M$ lattice sites, i.e: the Mott insulator
state (MI), superfluid state (SF), and a multisite coherent-state
approximation to the SF state (cf. Table~\ref{table1}).

\begin{table}
\begin{tabular}{|l|l|l|l|}\hline
& {\bf MI} & {\bf SF} & {\bf Coherent}\\ \hline \hline
$|\Psi\rangle$ & $\displaystyle\prod_{i=1}^M |n_i\rangle_i$ &
$\displaystyle\frac{1}{\sqrt{M^N N!}}(\sum_{i=1}^M
b_i^\dag)^N|0\rangle$ & $\displaystyle
e^{-\frac{N}{2}}\prod_{i=1}^M{e^{\sqrt{\frac{N}{M}}b_i^\dag}}
|0\rangle_i$
\\\hline $\langle\hat{n}_i^2\rangle$ &  $n^2$   & $n^2(1-1/N)+n$ &
$n^2+n$
\\\hline
$\left(\Delta n_i\right)^2$ &  0 & $n(1-1/M)$ & $n$
\\\hline
$\langle \hat{N}_K^2\rangle$ & $N_K^2$ & $N_K^2(1-1/N)+N_K$ &
$N_K^2+N_K$ \\\hline $(\Delta N_K)^2$ &  0 & $N_K(1-K/M)$ & $N_K$
\\\hline
$\langle\hat{n}_a\hat{n}_b\rangle$ & $n^2$ & $n^2(1-1/N)$ & $n^2$
\\\hline
$\langle\delta \hat{n}_a\delta \hat{n}_b\rangle$ & 0 & $-N/M^2$ & 0
\\\hline
\end{tabular}
\caption{\label{table1}Statistical quantities of typical atomic
states.}
\end{table}

The MI state represents a simple product of local Fock states at
each site with precisely $n_i$ atoms at a site $i$. As a
consequence, atom numbers at each site $\hat{n}_i$ (as well as the
number of atoms at $K$ sites $\hat{N}_K$) do not fluctuate, and
there is no quantum correlations between sites.

Similarly to the pair correlations, all two-, three-, and four-site
quantities in Eq.~(\ref{19}) factorize. From the light--scattering
point of view, this is the most classical atomic state, which
corresponds to periodically ordered pointlike atoms. We will further
consider the commensurate filling with $n_i=N/M$ atoms at each site,
neglecting possible random vacancies. This can be made if one has
some additional information that quantum fluctuations dominate over
other, thermal or technical, sources of noise.

The SF state corresponds to a BEC where each atom is in the zero
quasi-momentum Bloch--state of the lowest band and is equally
delocalized over all sites. Hence, the atom numbers at a given site
(and the number of atoms at $K<M$ sites) fluctuate. As a consequence
of the total atom number conservation, the numbers of particles at
two different sites $a\ne b$ are anticorrelated. All two-, three-,
and four-site quantities in Eq.~(\ref{19}) also do not factorize.

The expectation values in the SF state can be calculated using
normal ordering and the following relations:
\begin{subequations}
\begin{eqnarray}
b_i|\Psi_\text{SF}(N,M)\rangle=\sqrt{\frac{N}{M}}|\Psi_\text{SF}(N-1,M)\rangle,
\nonumber\\
\langle\Psi_\text{SF}|b_i^{\dag m}b_i^m|\Psi_\text{SF}\rangle =
\frac{N(N-1)...(N-m+1)}{M^m},\nonumber
\end{eqnarray}
\end{subequations}
where the first equation relates SFs with $N$ and $N-1$ atoms.

We will introduce another, coherent, quantum state, which is often
considered as an approximation to the SF state, and represents a
product of local coherent states at each site. In this approximate
state, the numbers of particles at a given site and at any $K\le M$
sites fluctuate. Moreover, the total number of particles at $M$
sites is also a fluctuating quantity, which is a disadvantage of
this approximation. Similarly to the MI state, correlations between
several different sites are absent. In the coherent state, one has
\begin{subequations}
\begin{eqnarray}
b_i|\Psi_\text{Coh}(N,M)\rangle=\sqrt{\frac{N}{M}}|\Psi_\text{Coh}(N,M)\rangle,
\nonumber\\
\langle\Psi_\text{Coh}|b_i^{\dag m}b_i^m|\Psi_\text{Coh}\rangle =
\frac{N^m}{M^m}.\nonumber
\end{eqnarray}
\end{subequations}

Comparing properties of the SF and coherent states in
Table~\ref{table1}, we can state that under the approximation
$N,M\rightarrow \infty$, but finite $N/M$, the coherent state is a
good approximation for local one-site quantities and correlations
between different sites. Moreover, if $K\ll M$, the SF expectation
values related to the nonlocal $\hat{N}_K$ operator are also well
approximated by corresponding quantities in the coherent state.
Nevertheless, if the number of sites $K$ is of the order of $M$, the
coherent-state approximation fails for those quantities.

One can prove even a more general statement for the functions
$\langle \hat{D}^*\hat{D} \rangle$ (\ref{13}) and $\langle
\hat{D}^{*2}\hat{D}^2\rangle$ (\ref{19}), which determine the
intensity and statistics of light and are the most important
quantities in this work. If the number of sites illuminated by
light, $K$, is much smaller than the total number of lattice sites
$M$, the coherent-state is a good approximation for calculating
characteristics of scattered light in the limit $N,M\rightarrow
\infty$, but finite $N/M$. If, in opposite, the number of sites
interacting with light is of the order of the total number of sites
in the lattice, this approximation, in general, gives wrong results.
As will be shown, it fails for light scattering in the directions of
diffraction maxima. The proof of the statement is based on the
consideration of the orders of sums in Eqs.~(\ref{13}), (\ref{19}),
which contain geometrical coefficients $A_i$ and are proportional to
the powers of $K$, whereas factors containing atom fluctuations have
powers of $M$ in denominators.

Thus, light scattering from the region of a SF optical lattice with
$K\ll M$ sites is equivalent to the light scattering from the atoms
in the coherent state (in absolute values both $K$ and $M$ can be
very large). Moreover, in the directions outside diffraction maxima,
the coherent-state approximation works well even in the case where
any number of sites is illuminated.

In the following, discussing all states, we will use the notations
$n=N/M$ for the atomic ``density'' (expectation value of the
particle number at each site) and $N_K=KN/M=nK$ for the expectation
value of the particle number at $K$ sites. These two parameters
fully characterize light scattering in the MI and coherent states,
while all three parameters $N$, $M$, and $K$ are necessary to
characterize scattering in the SF phase. For definitiveness, we will
discuss a case with large values of $N$, $M$, and $K$ where
difference between odd and even number of lattice sites vanishes.
Nevertheless, note that physical problems including BECs with large
atom number loaded into lattices with small site numbers are also of
great importance \cite{cennini2005ivn,albiez}. Results for this
case, can be obtained from expressions of this section and
Eqs.~(\ref{13}) and (\ref{19}).

\section{Example: 1D optical lattice in a transversally pumped cavity}

Before considering a general angular distribution of scattered
light, we would like to present the most striking prediction of our
model describing the difference between atomic quantum states,
observable by light scattering. Let us consider a configuration of
Fig.~\ref{fig1} where the pump (traveling or standing wave) is
orthogonal to the lattice ($\theta_0=0$), and the scattered light is
collected along the lattice axis ($\theta_1=\pi/2$) by a standing-
or traveling-wave cavity. This geometry coincides with the one
considered in the context of cavity cooling
\cite{vukics05,black2005clf,asboth2005soa} and lattices in optical
cavities \cite{maschler}. Atoms are assumed to be trapped at each
lattice site ($d=\lambda/2$) at field antinodes.

In this case, the operator $\hat{D}$ (\ref{12}) is reduced to
$\sum_{k=1}^K(-1)^{k+1}\hat{n}_k$, which, independently on an atomic
state, gives zero for the expectation value of the field amplitude
proportional to $\langle\hat{D}\rangle$ (here we assume even $K$).
This corresponds to the classical destructive interference between
atoms separated by $\lambda/2$. In contrast, the photon number in a
cavity $a_1^\dag a_1$ is proportional to
$\langle\hat{D}^*\hat{D}\rangle =(\langle\hat{n}^2\rangle - \langle
\hat{n}_a\hat{n}_b\rangle)K$ [cf. Eq.~(\ref{13c})], which is
determined by statistics of a particular state, and is equal to zero
for the MI state and to $N_K$ for the SF state.

Thus, atoms in a MI state scatter no photons into a cavity, while a
SF scatters number of photons proportional to the atom number:
\begin{eqnarray}
\quad \langle a_1\rangle_\text{MI}&=&\langle a_1\rangle_\text{SF}=0,
\quad \text{but} \nonumber\\
\langle a_1^\dag a_1\rangle_\text{MI}&=&0, \quad \langle a_1^\dag
a_1\rangle_\text{SF}=|C|^2N_K . \nonumber
\end{eqnarray}

Hence, already the mean photon number provides information about a
quantum state of ultracold atoms.

The photon number fluctuations $(\Delta n_\text{ph})^2$ (\ref{18})
are also different for various states. In the MI state, the variance
$(\Delta |D|^2)^2=\langle \hat{D}^{*2}\hat{D}^2\rangle -\langle
\hat{D}^*\hat{D}\rangle^2$ is zero, $(\Delta |D|^2)^2_\text{MI}=0$,
whereas in the SF state, Eq.~(\ref{19}) gives a very strong noise
$(\Delta |D|^2)^2_\text{SF}=2N_K^2$ (in highest order of $N_K$).

Nonlinear light-matter dynamics in a cavity can lead to a new
self-organized phase \cite{domokos2002cca,black2005clf} where all
atoms occupy only each second site leading to doubling of the
lattice period, $d=\lambda$. The operator $\hat{D}$ (\ref{12}) is
then reduced to $\sum_{k=1}^K\hat{n}_k=\hat{N}_K$. Thus, if the
final self-organized state is a MI with $d=\lambda$, the photon
number in a cavity is $\langle a_1^\dag
a_1\rangle_\text{Self-org}=|C|^2 N_K^2$, which is proportional to
the atom number squared and has a superradiant character. This
result coincides with the theory of self-organization with classical
center-of-mass motion \cite{domokos2002cca}.

\section{Results and Discussion}

In the following we will compare light scattering from atoms in the
following states: MI, SF with all sites illuminated ($K=M$ using the
notation SF$_M$), and partially illuminated SF under the
approximation $N,M\rightarrow \infty$, finite $n=N/M$, $K\ll M$,
which will be denoted as the ``coherent'' taking into account the
equivalence proved in Sec.~V. The results for the SF$_K$ state with
any $K$ can be obtained from the general Eqs.~(\ref{13}) and
(\ref{19}). We will restrict ourselves to the case of plane waves.
Distinguishing between atomic states using light modes with more
complicated spatial profiles can be analyzed by general expressions
of Sec. IV.

\subsection{Two traveling waves and discussion of essential physics}
\begin{figure}
\includegraphics{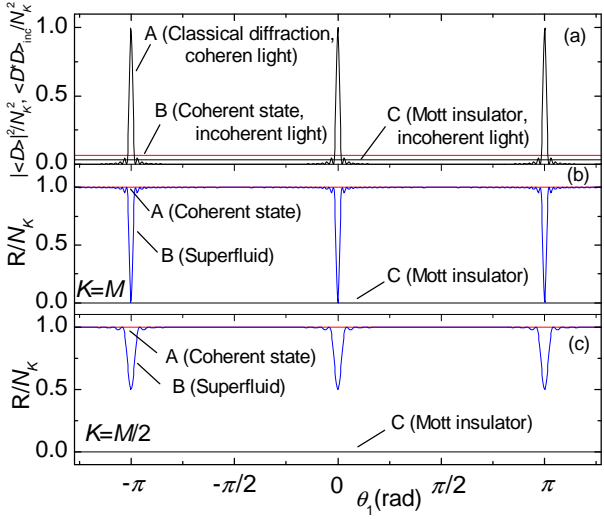}
\caption{\label{fig2}(Color online) Intensity angular distributions
for two traveling waves, the pump is transverse to the lattice
($\theta_0=0$). (a) Intensity of classical diffraction of coherent
light (curve A), isotropic intensity of incoherent light scattering,
Eq.~(\ref{14}), for coherent atomic state (line B) and MI state
(line C); (b) noise quantity, Eq.~(\ref{26}), for coherent atomic
state (constant value 1, line A), SF with all sites illuminated
$K=M$ (curve B), and MI (constant value 0, line C); (c) the same as
in (b) but for partially illuminated SF with $K=M/2$. $N=M=30$.}
\end{figure}

For two traveling waves, which can be free-space modes or fixed by
ring cavities, the geometrical coefficients (\ref{12}) are
$A_m=\exp(im \alpha_-)$ ($\alpha_-=k_{0x}d\sin\theta_0 -
k_{1x}d\sin\theta_1$), and Eq.~(\ref{13e}) for the noise quantity is
reduced to
\begin{eqnarray}\label{25}
R=\langle \delta\hat{n}_a\delta\hat{n}_b\rangle
\frac{\sin^2{(K\alpha_-/2})}{\sin^2{(\alpha_-/2})}
+(\langle\delta\hat{n}^2\rangle - \langle
\delta\hat{n}_a\delta\hat{n}_b\rangle)K,
\end{eqnarray}
where the first term has the angle dependence of classical
diffraction (\ref{11}), and the angle dependence in the second term
in Eq.~(\ref{13e}) is reduced to a constant (isotropic) one, $K$. In
the MI and coherent states, where pair correlations $\langle
\delta\hat{n}_a\delta\hat{n}_b\rangle$ are absent, the first term is
zero. In the MI state, on-site density fluctuations
$\langle\delta\hat{n}^2\rangle$ are also zero giving the zero value
of the noise quantity (\ref{25}), while in the coherent state, it is
the on-site fluctuations $\langle\delta\hat{n}^2\rangle=n$ that give
isotropic contribution to $R$. Thus, we have
\begin{subequations}\label{26}
\begin{eqnarray}
R_\text{MI}=0, \label{26a}\\
R_\text{Coh}=nK=N_K, \label{26b}\\
R_{\text{SF}_K}= -\frac{N}{M^2}
\frac{\sin^2{(K\alpha_-/2})}{\sin^2{(\alpha_-/2})}
+\frac{N}{M}K.\label{26c}
\end{eqnarray}
\end{subequations}

It is important to note, that in the SF state (\ref{26c}), even in a
large optical lattice with $N,M\rightarrow \infty$, very small pair
correlations $\langle \delta\hat{n}_a\delta\hat{n}_b\rangle=-N/M^2$
can give a significant angle-dependent contribution to the noise
quantity, which occurs near a diffraction maximum ($\alpha_-=2\pi l,
l=0,1,..$), where the geometrical factor is equal to $K^2$, and if
the number of the illuminated sites $K$ is of the order of $M$. This
demonstrates the importance of nonlocal correlations and invalidity
of the coherent-state approximation under those conditions. Outside
the diffraction maximum, where the geometrical factor is small, pair
correlations do not play any role and the coherent-state
approximation works well even for all sites illuminated.

Figure~\ref{fig2} shows several angle dependences of the scattered
light in the case of two traveling waves. As an example, in all
figures, we will consider atoms at each lattice sites providing
$d=\lambda_{0,1}/2$. In Fig.~\ref{fig2}(a), the angular distribution
of classical diffraction $|\langle D\rangle|^2$ (curve A) is shown.
In the case of $d=\lambda_{0,1}/2$ and the pump being orthogonal to
the lattice ($\theta_0=0$), only the zero-order diffraction maxima
at $\theta_1=0, \pi$ are possible in the classical picture.
Corresponding noise quantities $R$ for the coherent (constant lines
A) and SF$_K$ (curves B) states are displayed in Figs.~\ref{fig2}(b)
and \ref{fig2}(c) (in MI, the noise is zero, which is displayed by
lines C). According to Eq.~(\ref{26}), the intensity fluctuations
are isotropic for the coherent atomic state, while there is
suppression of intensity noise under scattering from the SF. The
suppression occurs in the regions of diffraction maxima. For all
sites illuminated, $K=M$ [cf. Fig.~\ref{fig2}(b)], the suppression
is total, while for $K=M/2$ it is only partial [cf.
Fig.~\ref{fig2}(c)]. Outside the maxima, the dependence for SF$_K$
is well approximated by that for the coherent state for any $K$.

It is important to underline, that in a broad range of angles, the
number of scattered photons from the SF (or coherent) state is
nonzero, even if the expectation value of the electromagnetic field
vanishes, which manifests the appearance of nonclassical
entanglement between the light and manybody atomic system. Moreover,
in contrast to MI state, atoms in SF state scatter photons at
angles, where the classical diffraction does not exist.

For example, in a simple configuration considered in Sec.~VI where
the pump is orthogonal to the lattices ($\theta_0=0$), and the
scattered light is collected by a cavity along the lattice axis
($\theta_1=\pi/2$), the atoms in the MI state scatter no photons as
in classical diffraction minimum. In contrast, atoms in the SF$_K$
state scatter the number of photons $a_1^\dag a_1=|C|^2\langle
\hat{D}^*\hat{D}\rangle=|C|^2 N_K$, proportional to the number of
the atoms illuminated [cf. Eq.~(\ref{26}) and Fig.~\ref{fig2}(b) at
the angle $\theta_1=\pi/2$].

For two traveling waves, the expression for $\hat{D}$ (\ref{12}), in
a diffraction maximum where all atoms radiate in phase with each
other and $\alpha_-=2\pi l$, is reduced to the operator $\hat{N}_K$.
Thus, the quantity $\langle \hat{D}\rangle=N_K=nK$ is the
expectation value of the atom number at $K$ sites and proportional
to the average atom number at a single site. The intensity of the
light scattered into a diffraction maximum is determined by
$\langle\hat{D}^*\hat{D} \rangle=\langle N_K^2\rangle$, while noise
$R = (\Delta N_K)^2$ gives the atom number variance at $K$ sites.
The latter statement corresponds to Figs.~\ref{fig2}(b) and
\ref{fig2}(c) displaying the total noise suppression in SF$_M$
state, where the total atom number at all sites $K=M$ does not
fluctuate, while for $K<M$, $N_K$ is a fluctuating quantity and the
noise suppression is only partial.

At the angle of a classical diffraction ``minimum'' (for $K\gg 1$
this is approximately valid for any angle outside narrow regions of
maxima), the expectation value of the field amplitude is zero, as
well as the first terms in Eqs.~(\ref{13c}), (\ref{13d}),
(\ref{13e}), and both the intensity $\langle\hat{D}^*\hat{D}
\rangle$ and noise $R$ are proportional to the quantity
$\langle\hat{n}^2\rangle - \langle \hat{n}_a\hat{n}_b\rangle$ giving
the difference between local and nonlocal fluctuations. For two
traveling waves, the coefficient of proportionality is isotropic and
equal to $K$ [cf. Eq.~(\ref{25})].

For scattering of incoherent light (\ref{14}), the intensity is
proportional to the local quantity $\langle \hat{n}^2\rangle$ and is
shown in Fig.~\ref{fig2}(a) for MI (curve C) and coherent, almost
the same as in SF, (curve B) states. This quantity can be also
obtained under coherent scattering of two traveling waves, if one
tunes the angles such that the geometrical factor of the first term
in Eq.~(\ref{25}) is equal to $K$. Practically, this variant is easy
to achieve only for a diffraction pattern with diffraction maxima,
which are not too narrow.

Hence, in an optical experiment, both global statistical quantities
related to $K\le M$ sites, local quantities reflecting statistics at
a single site, and pair correlations can be obtained. It is
important, that local statistics can be determined by global
measurements, i.e., an optical access to a single site is not
necessary.

Therefore, light scattering gives a possibility to distinguish
different quantum states of ultracold atoms. As demonstrated by
Eq.~(\ref{26}) and Fig.~\ref{fig2}, MI and SF$_M$ states are
distinguishable in diffraction ``minima'' and in incoherent light,
while they are indistinguishable (for traveling waves) in maxima,
because the total atom number contributing to the maximum does not
fluctuate. The SF$_M$ and coherent states can be distinguished in
diffraction maxima only. The MI and coherent states can be
distinguished in any angle of the scattering pattern.

Measurements of the noise quantity discussed or, alternatively,
related quantities for quadratures (\ref{16}) or photon number
variance (\ref{18}), give the values, which are different in orders
of the emitter number $N_K$ for different quantum states.
Nevertheless, for large $N_K$, there could be practical problems in
the subtraction of large values in a diffraction maximum to get the
noise contribution. In some papers, a similar problem even led to a
conclusion about state indistinguishability by intensity
measurements in BEC \cite{idziaszek,ciracPRL,ciracPRA} and, hence,
to a necessity to measure photon statistics. A rather involved
method to suppress the strong classical part of scattering using a
dark-state resonance in BEC was proposed in Ref.~\cite{mustPRA64}.
In contrast to homogeneous ensembles, in optical lattices, this
problem has a natural solution: measurements outside diffraction
maxima are free of the strong classical-like part and thus directly
reflect density fluctuations.

\subsection{Physical interpretation and role of the entanglement
between light and matter}

The classical analogy of the difference in light scattering from
different atomic states consists in various density fluctuations in
different states. In particular, classical density fluctuations
would also lead to impossibility of obtaining a perfect diffraction
minimum, where contributions from all sites should precisely cancel
each other.

Scattering at diffraction maxima can be treated as superradiant one,
since the intensity of the scattered light is proportional to the
number of phase-synchronized emitters squared $N_K^2$. In
diffraction minima, destructive interference leads to the total
(subradiant) suppression of coherent radiation for MI state; whereas
for SF$_K$ state, the intensity is nonzero and proportional to the
number of emitters $N_K$, which is analogous to the emission of
independent (non-phase-synchronized) atoms.

Nevertheless, the quantum treatment gives a deeper insight into the
problem.

The expression for the SF state in Table~\ref{table1} can be
rewritten in the following from:
\begin{eqnarray}
|\Psi_\text{SF}\rangle=\frac{1}{(\sqrt{M})^N}\sum_{q_i}
\sqrt{\frac{N!}{q_1!q_2!...q_M!}}|q_1,q_2,..q_M\rangle,\nonumber
\end{eqnarray}
where the sum is taken over all $q_i$ such that $\sum_i^M q_i=N$. It
shows that the SF state is a quantum superposition of all possible
multisite Fock states corresponding to all possible distributions of
$N$ atoms at $M$ lattice sites. Under the light--matter interaction,
the Fock states corresponding to different atomic distributions
become entangled to scattered light of different phases and
amplitudes~\cite{OptCom,Andras}.

For example, in a simple case of two atoms at two sites,
$|\Psi_\text{SF}\rangle =
1/2|2,0\rangle+1/\sqrt{2}|1,1\rangle+1/2|0,2\rangle$. In the example
configuration of Sec. VI, where the orthogonal pump illuminates
lattice sites separated by $\lambda/2$ (diffraction minimum), the
wave function of the whole light-matter system reads
\begin{eqnarray}
|\Psi_\text{matter-light}\rangle=1/2|2,0\rangle|\gamma\rangle+
1/\sqrt{2}|1,1\rangle|0\rangle \nonumber\\
+1/2|0,2\rangle|-\gamma\rangle.\nonumber
\end{eqnarray}

Here, if we assume that the distribution $|2,0\rangle$ is entangled
to the coherent state of light $|\gamma\rangle$, the distribution
$|0,2\rangle$ will be entangled to the similar light state with the
opposite phase $|-\gamma\rangle$, and the distribution $|1,1\rangle$
will be entangled to the vacuum field $|0\rangle$, because the
fields emitted by two atoms cancel each other.

In contrast to the classical case, light fields entangled to various
atomic distributions do not interfere with each other, which is due
to the orthogonality of the Fock states, providing a sort of
which-path information. This leads to a difference from the
classical (or MI with the only Fock state $|1,1\rangle$) case and
nonzero expectation value of the photon number even in the
diffraction minimum ($|\gamma|^2/2$ in the above example). The
absence of interference gives also an insight into the similarity of
scattering from the SF state to the scattering from independent
(non-phase-synchronized) atoms, where interference is also absent.

We would like to mention that light measurements considered here
correspond to the QND measurements~\cite{Brune} of atomic variables
observing light. Considering the Hamiltonian~(\ref{7}) with
neglected tunneling, it is straightforward to show that for the
``probe observable'' (i.e. measured light $a_1^\dag a_1$) and
``signal observable'' $\hat{D}$~(\ref{10b}), all four conditions of
the QND scheme summarized in Ref.~\cite{Brune} are fulfilled. Thus,
observing light one can determine various atomic functions
corresponding to the geometry-dependent operator
$\hat{D}$~(\ref{10b}) in a QND way. For example, the total atom
number or atom number at some lattice region can be nondestructively
measured in a diffraction maximum, while the difference between atom
numbers at odd and even sites can be nondestructively determined in
a diffraction minimum.

\subsection{Standing waves}
\begin{figure}
\includegraphics{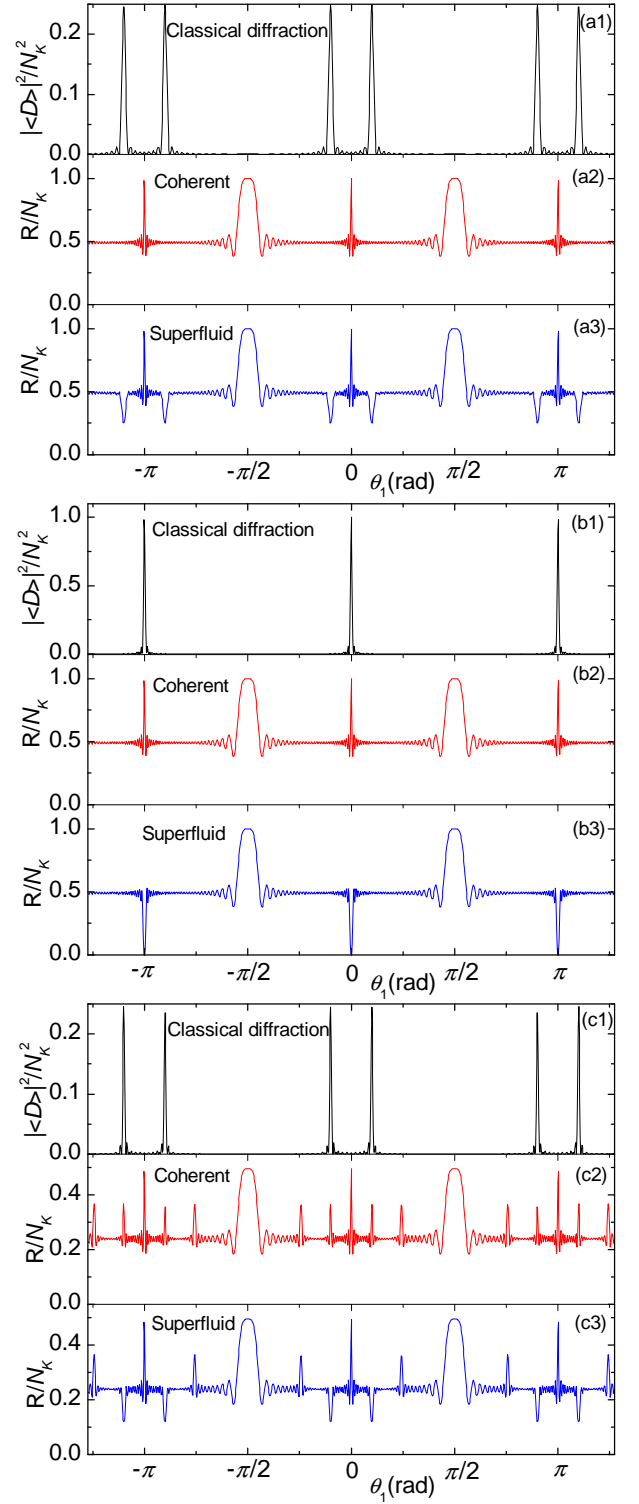}
\caption{\label{fig3}(Color online) Intensity angular distributions
for scattering into a standing-wave cavity. (a) Traveling-wave pump
at $\theta_0=0.1\pi$; (b) traveling or standing-wave pump at
$\theta_0=0$; (c) standing-wave pump at $\theta_0=0.1\pi$.
Intensities of classical diffraction are shown in Figs. (a1), (b1),
and (c1); noise quantities for coherent state are shown in Figs.
(a2), (b2), and (c2) and for SF in Figs. (a3), (b3), and (c3).
$N=M=K=30$.}
\end{figure}

\begin{figure}
\includegraphics{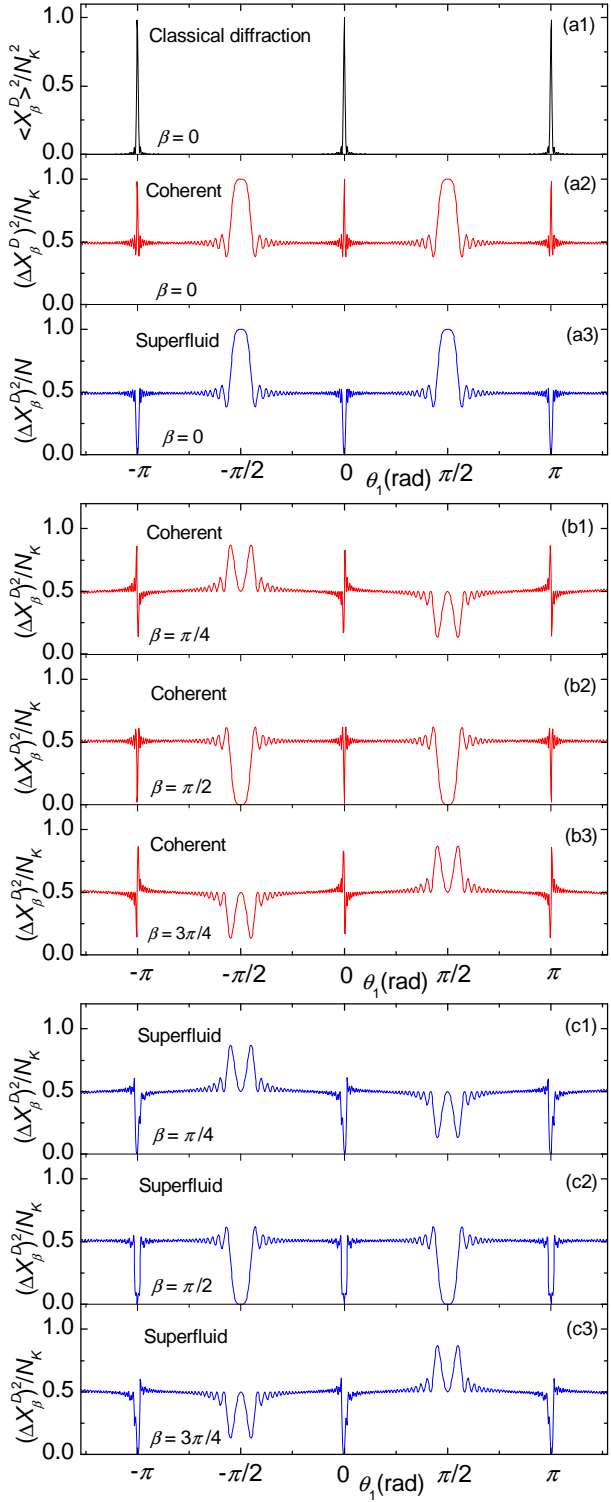}
\caption{\label{fig4}(Color online) Quadrature angular distributions
for two traveling waves. (a) Quadrature for classical diffraction
(a1), quadrature variance for coherent (a2) and SF (a3) states,
pump-homodyne phase difference $\beta=0$; (b) quadrature variance
for coherent state for $\beta=\pi/4$ (b1), $\beta=\pi/2$ (b2),
$\beta=3\pi/4$ (b3); (c) quadrature variance for SF for
$\beta=\pi/4$ (c1), $\beta=\pi/2$ (c2), $\beta=3\pi/4$ (c3).
$\theta_0=0$, $N=M=K=30$.}
\end{figure}

If at least one of the modes is a standing wave, the angle
dependence of the noise becomes richer. In an experiment, this
configuration corresponds to a case where the scattered light is
collected by a standing-wave cavity, whose axis can by tuned with
respect to the lattice axis \cite{maschler}. Except for the
appearance of new classical diffraction maxima represented by the
first terms in Eqs.~(\ref{13c}), (\ref{13d}), (\ref{13e}), which
depend on the phase parameters $\alpha_{\pm}=k_{0x}d\sin\theta_0 \pm
k_{1x}d\sin\theta_1$, the angle dependence of the second term is
also not an isotropic one, as it was for two traveling waves. This
second, ``noise,'' term includes a sum of the geometrical
coefficients squared, which is equivalent to the effective doubling
of the lattice period (or doubling of the light frequency) and leads
to the appearance of new spatial harmonics in the light angular
distribution. Such period doubling leads to the appearance of the
peaks in the noise distribution at the angles, where classical
diffraction does not exists.

In Fig.~\ref{fig3}(a), angular distributions of the scattered light
are shown for a traveling-wave pump, which is almost orthogonal to a
lattice ($\theta_0=0.1\pi$), while the probe is a standing wave.
Classical diffraction pattern [cf. Fig.~\ref{fig3}(a1)] is
determined by $|A|^2$ through the parameters $\alpha_\pm$ and shows
zero-order diffraction maxima in transmission ($\theta_1=\theta_0$
and its counterpart due to the presence of the standing-wave cavity
at $\theta_1=\pi+\theta_0$) and reflection ($\theta_1=\pi-\theta_0$
and the counterpart at $-\theta_0$). The intensity noise for atoms
in the coherent state [cf. Fig.~\ref{fig3}(a2)] is determined by
$\sum_{i=1}^K{|A_i|^2}$ through another parameter
$2\alpha_1=2k_{1x}d\sin\theta_1$ and has different characteristic
features at $\theta_1=0,\pi$, and $\pm \pi/2$. It is the latter
feature that corresponds to the effective frequency doubling and
appears at an angle, where classical diffraction has a minimum. In
the case of SF$_M$ state [cf. Fig.~\ref{fig3}(a3)], pair
correlations in Eqs.~(\ref{13d}) and (\ref{13e}) are nonzero, hence,
both geometrical factors contribute to the noise distribution, which
has the features at angles characteristic to both classical
scattering and the light noise of the coherent-state case. Outside
the characteristic features, the noise distribution is isotropic and
takes a nonzero value similar to the case of two traveling waves
[cf. Fig.~\ref{fig2}]. Figure~\ref{fig3}(b) shows a simpler
situation, where the pump is precisely orthogonal to the lattices
($\theta_0=0$).

In Fig.~\ref{fig3}(c), a situation similar to Fig.~\ref{fig3}(a) is
shown for the case where both the pump and probe are standing waves.
While classical diffraction still depends on the parameters
$\alpha_\pm$, the factor $\sum_{i=1}^K{|A_i|^2}$ determining the
intensity noise depends on four parameters
$2\alpha_{0,1}=2k_{0,1x}d\sin\theta_{0,1}$ and $2\alpha_{\pm}$.
Thus, in the light noise from a lattice in the coherent and SF
states, the features are placed at the positions of classical
zero-order diffraction maxima and the angles, which would correspond
to the classical scattering from a lattice with a doubled period
$d=\lambda$, where the appearance of first-order diffraction maxima
is possible. Similar to Fig.~\ref{fig3}(a), features at
$\theta_1=0,\pi$, and $\pi/2$ also exist. In the case $\theta_0=0$,
the angular distribution for two standing waves is identical to that
of one standing wave shown in Fig.~\ref{fig3}(b).

In the SF$_M$ state, there are two types of diffraction maxima. In
the first one, the noise can be completely suppressed due to the
total atom number conservation, similarly to the case of traveling
waves. This occurs, if the condition of the maximum is fulfilled for
both of two traveling waves forming a single standing wave [cf.
Fig.~\ref{fig3}(b)]. In the second type, even for $K=M$, only
partial noise suppression is possible, since only one of the
traveling waves is in a maximum, while another one, being in a
minimum, produces the noise [cf. Figs.~\ref{fig3}(a) and
\ref{fig3}(c)]. In contrast to two traveling modes, in the second
type of maxima, one can distinguish between SF$_M$ and MI states,
since MI produces no noise in any direction.

\subsection{Quadratures and photon statistics}

An analysis of the angular distribution of the quadrature variance
$(\Delta X_\beta^D)^2$ (\ref{16b}) shows, that even for two
traveling waves, new peaks due to effective period doubling appear
[see Fig.~\ref{fig4}(a)]. Additionally, the amplitude of noise
features can be varied by the phase difference between the pump and
homodyne beams $\beta$, which is shown in Figs.~\ref{fig4}(b) for
the coherent and in Fig.~\ref{fig4}(c) SF$_M$ states. In the
coherent state, all peaks are very sensitive to $\beta$. In the
SF$_M$ state, the noise suppression at diffraction maxima is
insensitive to variations of $\beta$, whereas other peaks are
$\beta$-dependent. The relation of $(\Delta X_\beta^D)^2$ to the
quadrature variance of the light field is given by Eq.~(\ref{15c}).

\begin{figure}
\includegraphics{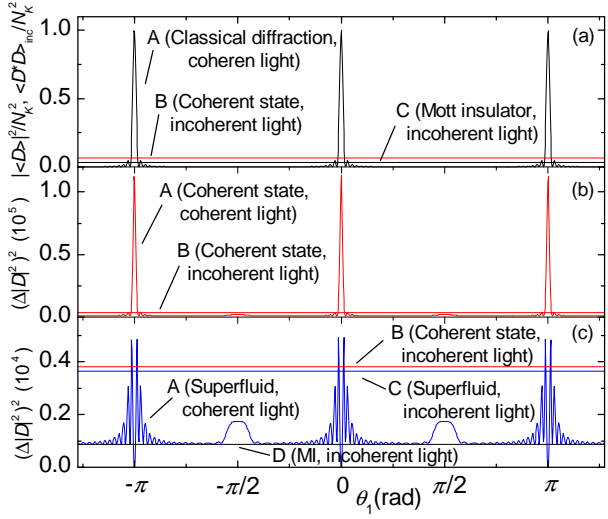}
\caption{\label{fig5}(Color online) Angular distributions of
photon-number variances for two traveling waves. (a) Intensity of
classical diffraction (curve A), isotropic intensity of incoherent
light scattering, Eq.~(\ref{14}), for coherent atomic state (line B)
and MI state (line C); (b) normal ordered photon-number variance for
coherent atomic state under scattering of coherent, Eq.~(\ref{27})
(curve A), and incoherent (line B) light; (c) normal ordered
photon-number variance for SF state under scattering of coherent,
Eq.~(\ref{27}) (curve A), and incoherent (line C) light, variance
for coherent (line B) and MI (curve D) states under scattering of
incoherent light. Normal ordered photon-number variance for MI state
under scattering of coherent light is zero for all angles.
$\theta_0=0$, $N=M=K=30$.}
\end{figure}

The angle dependence of the variance $(\Delta |D|^2)^2=\langle
\hat{D}^{*2}\hat{D}^2\rangle -\langle \hat{D}^*\hat{D}\rangle^2$,
which is proportional to the normal ordered photon-number variance
and determines the light statistics (\ref{18}), also shows
anisotropic features due to frequency doubling even for two
traveling waves (Fig.~\ref{fig5}). In this case, Eq.~(\ref{19}) is
reduced to
\begin{widetext}
\begin{eqnarray}\label{27}
\langle \hat{D}^{*2}\hat{D}^2\rangle =
\left(\frac{\sin(K\alpha_-/2)}{\sin(\alpha_-/2)}\right)^4 \langle
n_an_bn_cn_d\rangle +
2\left(\frac{\sin(K\alpha_-/2)}{\sin(\alpha_-/2)}\right)^3
\frac{\cos(K\alpha_-/2)}{\cos(\alpha_-/2)}(\langle
n_a^2n_bn_c\rangle-\langle n_an_bn_cn_d\rangle) \nonumber\\
-4\left(\frac{\sin(K\alpha_-/2)}{\sin(\alpha_-/2)}\right)^2
[(K-2)\langle n_an_bn_cn_d\rangle-(K-3)\langle n_a^2n_bn_c\rangle -
\langle n_a^3n_b\rangle] \nonumber\\
+ \left(\frac{\sin K\alpha_-}{\sin\alpha_-}\right)^2 (\langle
n_an_bn_cn_d\rangle-2\langle n_a^2n_bn_c\rangle + \langle
n_a^2n_b^2\rangle)+ 2K^2(\langle n_an_bn_cn_d\rangle -2\langle
n_a^2n_bn_c\rangle+\langle n_a^2n_b^2\rangle) \nonumber\\
+K(-6\langle n_an_bn_cn_d\rangle+12\langle n_a^2n_bn_c\rangle-
4\langle n_a^3n_b\rangle-3\langle n_a^2n_b^2\rangle+\langle
n^4\rangle),
\end{eqnarray}
\end{widetext}
where the first four terms has features at angles typical to
classical diffraction, the fourth term is also responsible for the
doubled-frequency feature, and the last two terms contribute to the
isotropic component.

For the coherent state, the light scattered into a diffraction
maximum displays a very strong noise (equal to $4N_K^3+6N_K^2+N_k$
because $\langle
\hat{D}^{*2}\hat{D}^2\rangle=N_K^4+6N_K^3+7N_K^2+N_k$ and $\langle
\hat{D}^*\hat{D}\rangle=N_K^2+N_K$), which is much stronger than the
isotropic component ($N_K^2$ in highest order of $N_K$) and the
features at $\theta_1=\pm \pi/2$ ($2N_K^2$ in highest order of
$N_K$) [Fig.~\ref{fig5}(b)]. In SF$_M$ state, the noise at maxima
can be suppressed, while at other angles, in highest order of $N_K$,
it is equal to that of the coherent state [Fig.\ref{fig5}(c)]. In MI
state, the variance $(\Delta |D|^2)^2$ is zero for all angles.
Conclusions about state distinguishing by measuring light statistics
are very similar to those drown from the intensity and amplitude
measurements, which have been discussed in Sec.~VIIA, including
scattering of incoherent light (see Fig.~\ref{fig5} and the
discussion of Fig.~\ref{fig2}).

In experiments, the nontrivial angle dependence of the noise can
help in the separation of the light noise reflecting atom statistics
from technical imperfections.

\section{Conclusions}

We studied off-resonant collective light scattering from ultracold
atoms trapped in an optical lattice. Measuring the light field
allows to characterize the quantum state of atoms in a
nondestructive way and in particular distinguish between different
atomic states. The scattered light differs in intensity, quadrature
variances, and photon statistics. A measurement of the intensity
angular distribution provides information about atom number
fluctuations in a finite lattice region, local quantum statistics at
single sites, and pair correlations. Note that even local statistics
can be determined by global measurements without an optical access
to particular sites. Alternatively to angle-resolved measurements,
variations of the mode wavelengths with respect to the wavelength of
a trapping beam can be considered.

Light scattering as a diagnostic tool has particular advantages for
optical lattices in contrast to scattering from a homogeneous BEC.
Here one has a natural way to suppress the strong classical
scattering background by looking at the directions of diffraction
minima. In these directions the expectation value of the field
amplitude vanishes while the intensity (photon number) is nonzero
and directly reflects quantum fluctuations. Furthermore, in an
optical lattice, the signal is sensitive not only to the periodic
density distribution, but also to the periodic density fluctuations,
giving an access to even very small nonlocal pair correlations.
These can be obtained by measuring light at diffraction maxima.

As the most striking example, we considered light scattering from a
1D lattice in a transversally pumped optical cavity as in a setup
involving collective cavity cooling
\cite{domokos2002cca,black2005clf,elsasser2004oba,slama2005psd}. The
number of photons scattered into the cavity is zero for the Mott
insulator phase but proportional to the atom number in the
superfluid phase. Both states have almost the same average density
but different quantum uncertainties. So the superfluid state is a
quantum superposition of different Fock states corresponding to all
possible distributions of $N$ atoms at $M$ sites. Under illumination
by a coherent light, various Fock states become entangled to
scattered light states with different amplitudes and phases. In
contrast to classical scattering, where the atoms are described by
c-number center-of-mass positions \cite{eichmann1993ysi}, for a
quantum description of the atomic motion the light field amplitudes
corresponding to different atomic distributions do not interfere.
This is due to the orthogonality of Fock states forming the
superfluid providing a sort of which way information.

In the example configuration, the cavity-field amplitude is
determined by the atom number operators $\hat{n}_i=b^\dag_ib_i$.
Hence, the expectation value of the field amplitude is sensitive to
the average density only. In contrast, the intracavity photon number
reflects the second moments of atom number operators (e.g.
density-density correlations), while photon statistics reflects the
forth moments.

Let's emphasize that other physical systems are possible, where the
light amplitude depends on the matter-field amplitudes $b_i$, while
the intensity is sensitive to density operators $\hat{n}_i$. The
latter situation is typical to configurations, where two or more
atomic subsystems exist and can interact with each other, in
particular, through light fields. The examples are matter-wave
superradiance and amplification, where two or more momentum states
of cold atoms were observed \cite{inouye,moore,pu,trifonov}, and
interaction between two BECs with different internal
\cite{zeng,ruostekoskiPRA55,searchberman,search} or motional
\cite{ruostekoskiPRA57} atomic states. In the framework of our
paper, matter-field amplitudes $b_i$ can also contribute to light
amplitudes $a_l$, if the tunneling between lattice sites is
important, which we have considered in the general model in Sec.~II.
In the rest of the paper, the lattice was assumed deep leading to
negligible tunneling.

In general, a variety of optical effects can be sensitive to a
quantum state of an ultracold matter, if the density operators enter
measurable quantities nonlinearly, such that the expectation value
of those quantities cannot be simply expressed through expectation
values of density operators. For instance, the $\chi^{(3)}$
nonlinearity \cite{yamamoto} and refractive index of a gas, where
nonlocal field effects are important \cite{dalibard}, were shown to
depend on atom statistics. In this paper, we focused on such
nonlinear quantities as intensity, quadrature variances, and photon
statistics of scattered light \cite{meystre}. Phase-sensitive and
spectral characteristics mentioned in Sec.~IV.D reflect the
dependence of the dispersion of a medium on the quantum state of
matter \cite{WeArxiv07,ICAP}. Moreover, such dispersion effects
(e.g. cavity-mode shift) will reflect atom statistics not only in
light intensity, but even in light amplitudes $\langle a_l\rangle$.

So far we have neglected the dynamic back action of scattered field
on atoms. This can be well justified in a deep lattice where the
momentum transfer is by far not enough to change the atomic
vibrational state as long as not too many photons are scattered.
Even without energy transfer the information one gets from the light
will induce measurement back-action. This should have intriguing
consequences for multiple consecutive measurements on the light
scattered from optical lattices.

\begin{acknowledgments}
This work was supported by the Austrian Science Fund FWF (grants
P17709 and S1512).
\end{acknowledgments}


\end{document}